\documentclass{aa}
\usepackage{graphics}
\usepackage{psfig}

\begin{document}

   \thesaurus{11.03.1;	 % galaxies: clusters: general
	      12.04.1;   % cosmology: dark matter
	      14.25.2}	 % X-rays: galaxies

\title{Reconstruction of radial temperature profiles of galaxy clusters}

\author{Yan-Jie Xue and Xiang-Ping Wu}

\offprints{Y.-J. Xue}
\mail{wxp@class1.bao.ac.cn}

\institute{Beijing Astronomical Observatory and 
National Astronomical Observatories,
Chinese Academy of Sciences, Beijing 100012, China}

   \date{Received 28 June, 2000; accepted 27 July, 2000}

   \titlerunning{Temperature Profiles of Clusters}
   \maketitle

\begin{abstract}
In this {\sl Letter} we present the radial temperature profiles of 
three X-ray clusters (A119, A2255 and A2256)
reconstructed from a combination of the X-ray
surface brightness measurements and the universal density profile
as the underlying dark matter distribution. Our algorithm is based
on the hydrostatic equilibrium for intracluster gas and the 
universality of the total baryon fraction within the virial radius.
The scaled temperature profiles of these three clusters appear to be
remarkably similar in shape, reflecting the 
underlying structural regularity, although they are inconsistent with
either isothermality or a significant decline with increasing
radius.  Nevertheless, we find a good agreement between our derived 
temperature profiles and the recent analysis of 11 clusters observed 
with BeppoSAX (Irwin \& Bregman 2000), which provides a useful clue to
resolving the temperature profile discrepancy raised recently in
literature. A comparison of our derived temperature
profiles with future spatially-resolved spectral measurements
may constitute a critical test for the standard model of
structure formation and the conventional scenario for dynamical
properties of clusters.
\end{abstract}


%
%  14.Sep.'90: Demo-Vs.
%________________________________________________________________

\section{Introduction}

The lack of robust constraints on the radial temperature profiles of hot gas 
contained within galaxy clusters 
is probably the major uncertainty in the present
determination of dynamical properties of clusters, which hinders 
clusters from acting as an ideal laboratory of testing theories of
formation and evolution of structures in the universe including 
a direct estimate of the cosmic mass density parameter $\Omega_{\rm{M}}$ 
by combining the baryon fraction measurement and the Big Bang
Nucleosynthesis. Indeed, previous studies have arrived at
conflicting results regarding the radial temperature gradients in
clusters. By analyzing 30 clusters observed with ASCA, 
Markevitch et al. (1998) claimed a significant temperature 
decline with radius quantified by a polytropic index 
of $1.2$-$1.3$ on the average. However, subsequent studies have soon 
raised doubt about the ubiquity and steepness of these temperature decline:
Irwin, Bregman \& Evrard (1999) carried out an analysis of the 
color profiles of the same clusters used by Markevitch et al. (1998)
but found an essentially flat temperature profile. 
Applying the spectral-imaging deconvolution method to 
a large sample of 106 ASCA clusters, White (2000) has showed that
90 percent of the temperature profiles are actually consistent 
with isothermality.  Further argument against the nonisothermality 
of intracluster gas has been put forward recently by Irwin \& Bregman (2000),
who reported the detection of a flat and even increasing temperature 
profile out to $\sim30\%$ of the viral radius for a sample of 
11 clusters observed with BeppoSAX.

Theoretically, it deserves an investigation into the possibility
of deriving the radial temperature profiles of intracluster gas
from the well-motivated physical mechanisms, incorporated with 
the X-ray imaging observations. This may provide a valuable clue to
resolving the above temperature profile discrepancy. There are 
two well-established facts on which we can rely today:
(1)The gravitational potential of a cluster is dominated by the 
dark matter distribution which can be described by the so-called universal
density profile, as suggested by a number of high-resolution simulations
(Navarro, Frenk \& White 1995 and hereafter NFW), 
although the innermost slope is still under debate. 
(2)The azimuthally-averaged X-ray surface brightness of a cluster 
is reliably measurable out to several or even $\sim10$ times as large 
as the X-ray core radius,
for which a good approximation is provided by the conventional $\beta$ model 
(Cavaliere \& Fusco-Femiano 1976). These two facts, along with
the hydrostatic equilibrium hypothesis and a reasonable choice of the boundary 
conditions, permit a unique determination of   
the gas temperature profile (Wu \& Chiueh 2000). On the other hand, 
a comparison of the theoretically expected temperature profile 
with the result from the X-ray spectroscopic measurement constitutes 
a critical test for the validity of the NFW profile and 
the hydrostatic equilibrium in clusters.

In this {\sl Letter}, we will attempt for the first time 
to derive the temperature profiles of 3 well-defined clusters with good X-ray
imaging observations extending to relatively large radii, based on 
the method developed by Wu \& Chiueh (2000). 
Our derived temperature profiles will be compared with 
the recent results of 11 clusters  observed with BeppoSAX 
(Irwin \& Bregman 2000).  We will examine the possible similarity in the
gas temperature profiles as a result of the underlying structural regularity
(e.g. Neumann \& Arnaud 1999). The implication of our results  
for the reported temperature profile discrepancy will be discussed. 
Throughout the {\sl Letter} we assume $H_0=50$ km s$^{-1}$ Mpc$^{-1}$ and a 
flat cosmological model with $\Omega_{\rm{M}}=0.3$ and $\Omega_{\Lambda}=0.7$.

\section{Theoretical expectations}

We briefly summarize the mathematical treatment of
the intracluster gas tracing the underlying dark matter distribution 
of clusters. First, if the X-ray surface brightness profile 
of a cluster can be well approximated by the conventional $\beta$ model 
(Cavaliere \& Fusco-Femiano 1976)
%1
\begin{equation}
S_{\rm{x}}(r)=S_0\left(1+\frac{r^2}{r_{\rm{c}}^2}\right)^{-3\beta+1/2},
\end{equation}
this would indicate (Cowie, Henriksen \& Mushotzky 1987),
%2
\begin{equation}
n_{\rm{e}}(r)T^{1/4}(r)=n_{\rm{e}0}T_0^{1/4}
\left(1+\frac{r^2}{r_{\rm{c}}^2}\right)^{-3\beta/2}
\end{equation}
for an optically-thin, thermal bremsstrahlung emission, where 
$n_{\rm{e}}$ and $T$ are the electron number density and temperature,
respectively. The central electron number
density $n_{\rm{e}0}$, temperature $T_0$ and X-ray surface brightness $S_0$ 
are connected by 
%3
\begin{equation}
n_{\rm{e}0}^2=\frac{4\pi^{1/2}}{\alpha(T_0)\mu_{\rm{e}} g}
        \frac{\Gamma(3\beta)}{\Gamma(3\beta-1/2)}
        \frac{S_0(1+z)^4}{r_{\rm{c}}},
\end{equation}
where $\alpha(T_0)=(2^4e^6/3m_{\rm{e}}\hbar c^2)(2\pi kT_0/3m_{\rm{e}}c^2)^
{1/2}$,
$\mu_{\rm{e}}=2/(1+X)$ with $X$ being the primordial hydrogen mass fraction,
$g\approx1.2$ is the average Gaunt factor, and $z$ is the cluster redshift.
The total mass in gas within $r$ is simply
%4
\begin{equation}
M_{\rm{gas}}(r)=4\pi\mu_{\rm{e}} m_p n_{\rm{e}0} \int 
\left(\frac{T_0}{T}\right)^{1/4}
\left(1+\frac{r^2}{r_{\rm{c}}^2}\right)^{-3\beta/2} r^2 dr.
\end{equation}
Secondly, if the intracluster gas is in hydrostatic equilibrium with the
underlying dark matter distribution, we have
%5
\begin{equation}
\frac{GM_{\rm{DM}}(r)}{r^2}=
-\frac{1}{\mu m_{\rm{p}} n_{\rm{e}}}   \frac{d(n_{\rm{e}}kT)}{dr}.
\end{equation}
where $\mu=0.585$ is the average molecular weight. For NFW profile 
%6
\begin{equation}
M_{\rm{DM}}(r)=4\pi\rho_{\rm{s}} r_{\rm{s}}^3\left
[\ln\left(1+\frac{r}{r_{\rm{s}}}\right)-
          \frac{r}{r+r_{\rm{s}}}\right].
\end{equation}
Here we have neglected the self-gravity of the gas. Using the normalized 
gas temperature $\tilde{T}(r)\equiv T(r)/T_0$ and 
the volume-averaged baryon fraction $f_{\rm{b}}(r)\equiv 
M_{\rm{gas}}(r)/M_{\rm{DM}}(r)$
as the two variables, we obtain the following two first-order 
differential equations
%7,8
\begin{eqnarray}
\frac{d\tilde{T}}{dx}=\frac{4\beta x \tilde{T}}{x^2+a^2}-
 \frac{4\alpha_0}{3x^2}\left[\ln(1+x)-\frac{x}{1+x}\right];\\
\frac{df_{\rm{b}}}{dx}=\frac{b\tilde{T}^{-1/4}
                     (1+x^2/a^2)^{-3\beta/2}x^2
                -f_{\rm{b}} x/(1+x)^2}
                {\ln(1+x)-x/(1+x)},
\end{eqnarray}
where $x=r/r_{\rm{s}}$, $a=r_{\rm{c}}/r_{\rm{s}}$, $b=\mu_{\rm{e}}
n_{\rm{e}0}m_{\rm{p}}/\rho_{\rm{s}}$ and
$\alpha_0=4\pi G \mu m_{\rm{p}} \rho_{\rm{s}} r_{\rm{s}}^2/kT_0$.  
The first equation
can be straightforwardly solved with $\tilde{T}(0)=1$:
%9
\begin{equation}
\tilde{T}(x)=\left(1+\frac{x^2}{a^2}\right)^{2\beta}
             \left[1-\frac{4\alpha_0}{3}\int_0^x 
               \frac{\ln(1+x)-x/(1+x)}{x^2(1+x^2/a^2)^{2\beta}}dx\right].
\end{equation}
In order to solve the second equation and determine the free parameters
$a$, $b$ and $\alpha_0$, we use the following boundary conditions
%10,11
\begin{eqnarray}
f_{\rm{b}}(r_{\rm vir})=f_{\rm{b,BBN}};\\
\frac{df_{\rm{b}}}{dx} \left|_{x=r_{\rm{vir}}/r_{\rm{s}}}=0. \right.
\end{eqnarray}
Namely, we demand that  the baryon fraction should 
asymptotically match the universal value of $f_{\rm{b,BBN}}$ 
at the virial radius $r_{\rm{vir}}$ defined by
%12
\begin{equation}
M_{\rm{DM}}(r_{\rm{vir}})=\frac{4\pi}{3}r_{\rm{vir}}^3 \Delta_{\rm{c}} 
\rho_{\rm{crit}},
\end{equation}
where $\Delta_{\rm{c}}$ represents the overdensity of 
dark matter with respect to
the average background value $\rho_{\rm{crit}}$, for which we take 
$\Delta_{\rm{c}}=178\Omega^{0.45}_{\rm{M}}(z)$  and 
$\Omega_{\rm{M}}(z)=\Omega_{\rm{M}}(1+z)/\{1+z\Omega_{\rm{M}}+
[(1+z)^{-2}-1]\Omega_{\Lambda}\}$.
We now come to the free parameters involved in eqs.(8) and (9). 
With the X-ray imaging observation, we can obtain the best-fit 
values of $\beta$, $r_{\rm{c}}$ and $S_0$. 
If, on the other hand, the X-ray spectroscopic measurement can set 
a useful constraint on the central temperature $T_0$, we will be able to 
derive the central electron density from  eq.(3).  As a result, there
are only two free parameters in the above equations: $\rho_{\rm{s}}$ 
(or equivalently $\delta_{\rm{c}}=\rho_{\rm{s}}/\rho_{\rm{crit}}$) and 
$r_{\rm{s}}$. These two parameters can be
fixed during the numerical searches for the solution of eqs.(8) and (9)
using the boundary conditions eqs.(10) and (11).  
This will allow us to work out simultaneously the radial profiles of gas 
density and temperature, and fix
the dark matter (NFW) profile of the cluster characterized by 
$\rho_{\rm{s}}$ and $r_{\rm{s}}$.

\section{Application to X-ray clusters}

Since the reconstruction of gas temperature profile is sensitive 
to the initial input of $S_{\rm{x}}$ especially the $\beta$ parameter, 
whether or not we can reliably derive the temperature profile 
depends critically on the goodness of the single $\beta$ model fit 
to the X-ray surface brightness profile. 
We thus restrict ourselves to  the X-ray 
flux-limited sample of 45 clusters published recently by 
Mohr, Mathiesen \& Evrard (1999), in which there are sufficiently large 
data points to set robust constraints on the $\beta$ model fit.
The inclusion of a cluster is based on the following two criteria:
(1)The X-ray surface brightness profile can be well fitted by 
a single $\beta$ model with $0.8\leq \chi_{\rm{{\nu}}}^2\leq1.25$;
(2)The maximum extension ($r_{\rm{m}}$) of the X-ray observed surface 
brightness profile should be large enough to guarantee
the validity of the $\beta$ model at the outermost regions of clusters. Here
we set $r_{\rm{m}}\geq1.5$ Mpc. Unfortunately, it turns out that 
there are only three clusters which meet our criteria 
(Table 1): A119, A2255 and A2256.  In fact, our first criterion 
implies that the effect of cooling flows in the central regions of clusters
should be negligibly small. This explains the fact that the three selected
clusters all have large core radii. Note that the presence of cooling flows
may lead to the failure of a single $\beta$ model fit to the X-ray  
surface brightness profiles. In other words, our method cannot be applied to 
the clusters with strong cooling flows.  While the X-ray imaging data  
of the clusters can be somewhat accurately acquired, the present X-ray spectral
measurements have yielded the emission-weighted temperatures rather than 
the central values $T_0$ appearing in the $\alpha_0$ parameter.
Therefore, we have to use the emission-weighted temperature 
as a first approximation of $T_0$. Alternatively, we adopt 
the universal baryon fraction $f_{\rm{b,BBN}}=1/6$ to 
reconcile our cosmological
model of $\Omega_{\rm{M}}=0.3$ (for $\Omega_{\rm{b}}=0.05$).

 \begin{table}
 \vskip 0.2truein
 \scriptsize
 \begin{center}
 \caption{Cluster Sample}
 \begin{tabular}{llll}
 \hline
  & & &  \\ 
 cluster & A119 & A2255 & A2256 \\
   & & &  \\ 
  $z$       & 0.0438  & 0.0808  & 0.0581 \\
  $T_0$ (keV) & 5.80    & 7.30    & 7.51 \\
  $S_0^*$ & 1.18 & 1.68 & 4.41 \\
 $\beta$   & 0.662 &  0.792 & 0.828 \\
 $r_{\rm{c}}$ (Mpc) & 0.494  & 0.608     &  0.500 \\
 $n_{\rm{e}0}$ ($10^{-3}$cm$^{-3}$) & $1.37 $ & 1.67 & 2.94 \\
 $b$  &  4.15 & 3.48 & 3.31 \\
$\alpha_0$ & 17.80(12.30)$^+$ & 14.66(24.80)$^+$ & 13.15(13.21)$^+$\\
$\delta_{\rm{c}}$ & 130(490)$^+$     & 184(230)$^+$     & 345(1220)$^+$ \\
$r_{\rm{s}}$ (Mpc) & 5.77(2.59)$^+$ & 4.82(5.99)$^+$    & 3.43(2.03)$^+$\\
  &  &  & \\
 \hline
 \end{tabular}
 \end{center}

\parbox{8.5in}{$^*$In units of $10^{-13}$ erg s$^{-1}$ cm$^{-2}$ 
               arcmin$^{-2}$ for energy band 0.5-2.0 keV;}
\parbox{8.5in}{$^+$The result for an isothermal gas distribution.}
  \end{table}

Using the available X-ray data of the three clusters from Mohr et al. (1999),
we have performed numerical searches for the solutions of eqs.(8) and (9)
by iterations until the boundary conditions eqs.(10) and (11)
are satisfied. The resulting parameters $\alpha_0$,  
$\delta_{\rm{c}}$ and $r_{\rm{s}}$ are summarized in Table 1, together with a 
comparison with the corresponding values for an isothermal
gas distribution estimated in previous work (Wu \& Xue 2000).
Most importantly, such a procedure enables us to completely fix
the radial profiles of gas density, 
temperature and baryon fraction for the three clusters. 
Here we have no intention to illustrate the radial variations of 
$n_{\rm{e}}(r)$ and $f_{\rm{b}}(r)$, 
which essentially follow the theoretical expectations
(Wu \& Chiueh 2000). Rather, we display in Fig.1 the radial  
profiles of the emission-weighted temperatures for 
the three clusters constructed from our algorithm. 
Surprisingly, none of the temperature profiles of these three clusters
are consistent with the conventional speculations, and a visual examination
of Fig.1 reveals that they are neither characterized by isothermality nor 
represented simply by the polytropic equation of state. 
Nevertheless, these temperature profiles indeed demonstrate a similar radial 
variation, reflecting probably the underlying structural regularity.
Basically, the radial variation of the gas temperature resembles a
distorted `S' in shape: 
There exist two turnover points roughly at $0.1r_{\rm{vir}}$ 
and $0.4r_{\rm{vir}}$, respectively, where $dT/dr=0$, which separate 
the temperature curve $T(r)$ into three parts -- a decreasing $T(r)$ 
with radius inside the cluster core of $\sim0.1r_{\rm{vir}}$,  following
a slightly increasing $T(r)$ until $\sim0.4r_{\rm{vir}}$, 
and finally a moderately
decreasing $T(r)$ out to the virial radius. Overall, the absolute values of 
the gas temperature do not demonstrate a dramatic change within 
clusters.

%\placefigure{fig1}
    \begin{figure} 
	\psfig{figure=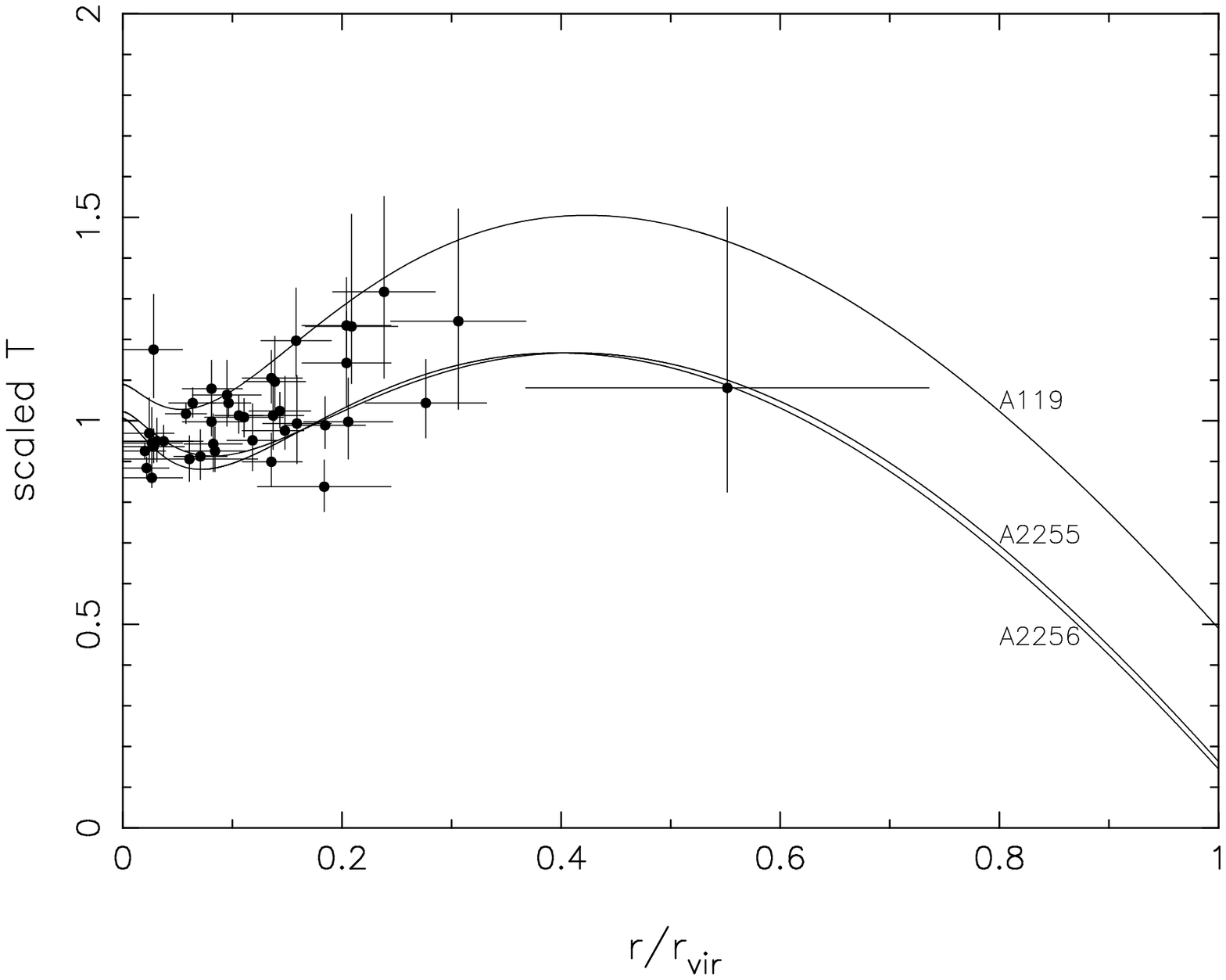,width=88mm,bbllx=80pt,bblly=120pt,bburx=540pt,bbury=650pt,clip=,angle=270}
	\caption{A comparison of the derived radial temperature profiles
of three clusters (A119, A2255 and A2256) with the results of 11
clusters observed with BeppoSAX (Irwin \& Bregman 2000).
The observed data are normalized  by the mean temperature for each
cluster, while the derived temperature curves are scaled by $1.32T_0$
for comparison. The horizontal axis is in units of the virial radius. }
\label{fig1}
   \end{figure}

The azimuthally-averaged radial temperature profiles of 11 clusters 
derived by Irwin \& Bregman  (2000) from an analysis of the BeppoSAX data
are superimposed on Fig.1. It appears that our derived temperature profiles 
are in good agreement with their observed ones over entire radius range. 
In fact, the significant temperature discrepancy raised in different 
X-ray spectral measurements occurs in the inner parts of clusters.
In the outer regions, it seems that many observations have provided   
a moderately decreasing temperature profile, 
which is essentially consistent with our theoretical predictions.  
Alternatively, our result is also compatible with the gas temperature 
distribution at large radii revealed by numerical simulations that 
demonstrate a temperature
decline of $\sim30\%$ of the central value at the virial radius
(Frenk et al. 1999).

\section{Discussion and conclusions}

In the absence of the detailed information on the radial temperature
profiles of clusters from X-ray spectroscopic measurements, we have 
made an attempt to derive the gas temperature profiles by combining
the X-ray surface brightness measurements and
the NFW profile as the underlying dark matter distribution of 
clusters. This has become possible when the intracluster gas is required  
to satisfy the hydrostatic equilibrium and the volume-averaged baryon 
fraction within the viral radius is required to asymptotically match 
the universal value. Consequently, we have obtained semi-analytically  
the temperature profiles of three clusters selected carefully from 
the ROSAT observed cluster sample. 
These derived temperature profiles are consistent with 
the new observations of 11 BeppoSAX clusters (Irwin \& Bregman 2000) 
and other measurements made at large cluster radii 
(e.g. Markevitch et al. 1998) as well as the result given by
numerical simulations (e.g. Frenk et al. 1999).

Regardless of the small sample, the three clusters exhibit a 
temperature profile similar in shape when the length scales are normalized to
their virial radii,  perhaps indicative of the underlying structural 
regularity.  
The present study provides a helpful clue to resolving the temperature
profile discrepancy: It is very likely that the lack of the high-quality 
data of the spatially resolved spectral observations 
would yield an emission-weighted temperature roughly close to isothermality 
within $\sim80\%$ of the virial radius, which does not exclude
the possibility that a slightly increasing 
temperature profile may be marginally detectable in the range of  
$0.1r_{\rm{vir}}<r<0.4r_{\rm{vir}}$. This explains the recent observations of 
Irwin \& Bregman (2000) and other studies (e.g. Kikuchi et al. 1999;
White 2000; etc.). However, our model does not predict
the flat temperature profile toward the inner regions of clusters 
as reported particularly by Markevitch et al. (1998), although 
a moderately decreasing temperature
profile will ultimately take place in the outer clusters ($r>0.4r_{\rm{vir}}$).

A conclusive test for the universality of our derived temperature profiles 
can be provided by future X-ray spectroscopic measurements.  Indeed, 
it will be useful to apply the present method to other 
X-ray clusters with good X-ray surface brightness profiles 
measured to large radii and  high-quality data of the spatially-resolved 
spectral observations at least within the central regions.
This may allow us to further justify our model and include the measurement 
uncertainties which have been neglected in the present study.
The inconsistency of the predicted temperature profiles
with the X-ray spectroscopic results will challenge the prevailing
models of structure formations as well as the conventional scenario of
cluster dynamics such as the hydrostatic equilibrium.
Finally, we should point out that our proposed method to obtain
the temperature profiles of clusters can be 
significantly contaminated by nongravitational heating
processes especially from the supernova-driven protogalactic winds.
Recall that the asymptotic tendency of the derived temperature
profiles at large radii depends sensitively on the $\beta$
parameter, while the energy injection of supernovae and active galaxies 
into the intracluster gas will result in a shallower 
X-ray surface brightness distribution (David et al. 1990; Ponman, Cannon \&
Navarro 1999; Llyod-Davies, Ponman \& Cannon 2000). Without
correction to this effect the theoretically predicted temperature profiles
may rise too rapidly at large radii.  
Note that at large radii the NFW mass profile 
diverges logarithmically with $r$, which differs significantly from
the variation of the gas mass profile (roughly $M_{\rm{gas}}\propto r$) 
expected from the assumption of isothermality.  
For a cluster with smaller $\beta$, $r_{\rm{c}}$ and $r_{\rm{s}}$, 
an increasing temperature
profile near virial radius is thus required to maintain the 
universality of the cluster baryon fraction. 
Therefore, a robust, theoretical determination of
the temperature profiles of clusters  should also allow 
nongravitational heating processes to be included.

\begin{acknowledgements}
We gratefully acknowledge Tzihong Chiueh for useful 
discussion and comments, and an anonymous referee for valuable suggestions.
This work was supported by 
the National Science Foundation of China, under Grant 19725311.
\end{acknowledgements}


\begin{thebibliography}{}

\bibitem{}Cavaliere A.,  Fusco-Femiano R., 1976, A\&A, 49, 137
\bibitem{}Cowie L. L., Henriksen M., Mushotzky R. F., 1987, 
            ApJ, 317, 593
\bibitem{}David L. P., Arnaud K. A., Forman W., Jones C., 
                1990, ApJ, 356, 32 
\bibitem{}Frenk C. S., et al., 1999, ApJ, 525, 554
\bibitem{}Irwin J. A., Bregman J. N., 2000, ApJ, in press
\bibitem{}Irwin J. A., Bregman J. N.,  Evrard A. E., 1999, 
                ApJ, 519, 518
\bibitem{}Kikuchi K., et al., 1999, PASJ, 51, 301
\bibitem{}Lloyd-Davies E. J., Ponman T. J., Cannon D. B., 
                2000, MNRAS, in press
\bibitem{}Markevitch M., Vikhlinin A., Forman W. R.,  
	     Sarazin C. L., 1998, ApJ, 527, 545  
\bibitem{}Mohr J. J., Mathiesen B., Evrard A. E., 1999, ApJ, 
	    517, 627 
\bibitem{}Navarro J. F., Frenk C. S., White S. D. M., 1995, MNRAS, 
            275, 720 (NFW)
\bibitem{}Neumann D. M., Arnaud M., 1999, A\&A, 348, 711
\bibitem{}Ponman T. J., Cannon D. B., Navarro J. F., 1999, 
                Nature, 397, 135
\bibitem{}White D. A., 2000, MNRAS, 312, 663
\bibitem{}Wu X.-P.,  Chiueh T., 2000, ApJ, in press
\bibitem{}Wu X.-P,  Xue, Y.-J.,  2000, ApJ, 529, L5 

\end{thebibliography}
\end{document}